\documentclass[11pt]{article}

\usepackage{graphicx}
\title{Accurate kinetic energy evaluation in electronic
 structure calculations with localised 
functions on real space grids}
\author{Chris-Kriton Skylaris\thanks{Corresponding author. 
Email: cks22@phy.cam.ac.uk.  Fax: +44 (0)1223 337356.}, Arash A. Mostofi, Peter D. Haynes
\and Chris J. Pickard and Mike C. Payne \\ \\
Theory of Condensed Matter, Cavendish Laboratory, \\ 
Madingley Road, Cambridge CB3 0HE, UK}
\date{  }
\begin{document}
\maketitle
\begin{abstract}
We present a method for calculating the kinetic energy  of localised 
functions represented on a regular real space grid.
This method uses fast Fourier transforms applied to restricted
regions commensurate with the simulation cell and is applicable
to grids of any symmetry. In the limit of large systems 
it scales linearly with system size. Comparison with the finite difference 
approach shows that our method offers significant improvements 
in accuracy without loss of efficiency. 
\end{abstract}

{ \noindent
\hspace{-1.5em} \textbf{Published as Computer Physics 
Communications \underline{140} (2001) 315-322}  }

\section{Introduction}

Density functional theory (DFT) combined with the pseudopotential method has been 
established as an important theoretical tool for studying a wide range of problems in 
condensed matter physics \cite{PAYNE_REVIEW}. However, the computational cost of 
performing a total-energy calculation on a system scales asymptotically as the cube 
of the system size. Consequently, plane-wave pseudopotential 
DFT can only be used to study systems of up to about one hundred 
atoms on a single workstation and up to a few hundred atoms on 
parallel supercomputers. As a result there has been considerable 
recent effort in the development of methods whose computational cost 
scales linearly with system size \cite{GOEDECKER-REVIEW}.

A common feature of many of the linear-scaling strategies is the 
expansion of the single-particle density matrix in terms of a set 
of localised functions. We refer to these functions 
as `support functions' \cite{GILLAN95}. A support function is required 
to be non-zero only within a spherical region, which we refer to 
as a `support region',  centred on an atomic position. Here we consider 
a representation of the support functions in terms of a regular 
real space grid, which constitutes our basis set. If the set 
of support functions is $\{\phi_\alpha\}$, the single-particle 
density matrix is expressed as 
\begin{equation}
\rho(\mathbf{r}',\mathbf{r})=\sum_{\alpha,\beta}\phi_\alpha(\mathbf{r}')
K^{\alpha \beta} \phi^{*}_\beta(\mathbf{r}) \ ,
\end{equation}  
where $K^{\alpha\beta}$ is the matrix representation of the 
density matrix in terms of the duals of the support functions. 
In general the support function set is not orthonormal.

Real space methods have the advantage that they provide a clear spatial 
partitioning of all quantities encountered in a density functional 
calculation, a property that is ideal for code parallelisation. As 
a result, this approach has gained popularity in recent years 
and a  number of such density functional calculations 
have been reported by different 
authors \cite{CHELIKOWSKY2,BRIGGS-BERNHOLC,MARTIN-MULTI}. These 
approaches use  finite difference (FD) methods \cite{ACTA_NUMERICA_94} 
for the calculation of the kinetic energy. In terms of the 
support functions the kinetic energy is
\begin{equation} 
E_T[\rho]=\sum_{\alpha, \beta} T_{\alpha \beta } K^{\beta \alpha} \ , \label{MATRIX_KINETIC}
\end{equation}
where  $T_{\alpha \beta}$ denotes kinetic energy matrix 
elements between support functions, given in Hartree atomic units by
\begin{equation}
T_{\alpha \beta}= -\frac{1}{2} \int \phi^*_{\alpha}(\mathbf{r}) 
\nabla^2 \phi_{\beta}(\mathbf{r}) \mathrm{d} \mathbf{r}  \label{TAB}  \ .
\end{equation}
The evaluation of the kinetic energy matrix elements requires 
the action of the Laplacian operator on the support functions. Here 
we will show that in the case of localised support 
functions, fast Fourier transform (FFT) methods can be adapted 
for the application of the Laplacian, providing an algorithm 
with essentially the same computational cost as FD but 
with  higher accuracy and also ready
applicability to any grid symmetry. 

In the following two sections we present 
the FD method and our new FFT-based 
method and compare them both in theory and in practice.

\section{Theory}

For functions represented as values on a regular 
grid, integrals like the one of equation (\ref{TAB}) can 
be calculated, or rather approximated to increasing 
accuracy, by a sum over grid points, as long as 
the value of the integrand is known at every grid point:
\begin{equation}
T_{\alpha \beta} \simeq -\frac{1}{2} w  \sum_{\mathbf{r}_i} 
\phi^*_\alpha(\mathbf{r}_i)
\hat{T} \phi_\beta(\mathbf{r}_i )   \ ,  \label{REAL_SUM}
\end{equation}
where $\hat{T}$ is the Laplacian operator in the discrete 
representation, $w$ is the volume per grid point, and 
the sum formally goes over all the grid points in the simulation cell.

\subsection{Finite Differences}

The most  straightforward approach to the evaluation of the 
Laplacian operator applied to a function at every grid point 
is to approximate the second derivative by finite differences of 
increasing order of accuracy \cite{ACTA_NUMERICA_94}. For 
example, the $\partial^2 \phi / \partial x^2$ part of the 
Laplacian on a grid of orthorhombic symmetry is
\begin{equation}
  \frac{\partial^2 \phi}{\partial x^2} (x_i, y_j, z_k) 
\simeq \frac{1}{h_{x}^2}\sum_{n=-A/2}^{A/2} C_n^{(A)} 
\phi(x_i+nh_{x}, y_j, z_k) + O(h_{x}^{A}) \ ,  \label{LAPLACIAN_FD}
\end{equation}
where $h_{x}$ is the grid spacing in the $x$-direction, $A$ is 
the order of accuracy and is an even integer, and the 
weights $C_n^{(A)}$  are even with respect 
to $n$, i.e. $C_n^{(A)}=C_{-n}^{(A)}$. This equation is 
exact when $\phi$ is a polynomial of degree less than or 
equal to $A$. The leading contribution to the error is 
of order $h_{x}^{A}$. The full Laplacian operator for a 
single grid point in three dimensions consists of a 
sum of $(3A+1)$ terms.

In principle, for well behaved functions, the second 
order form of equation (\ref{LAPLACIAN_FD}) should converge 
to the exact Laplacian as $h \rightarrow 0$. Therefore 
to increase the accuracy of a calculation one would 
need to proceed to smaller grid spacings. However, in 
most cases of interest, this is computationally 
undesirable and instead, formulae of increasing order 
are used to improve the accuracy at an affordable 
cost \cite{BECK-REVIEW}. Chelikowsky 
et al. \cite{CHELIKOWSKY1}, in their finite difference 
pseudopotential method, have tested the finite 
difference expression for up to $A=18$ on calculations 
of a variety of diatomic molecules and have 
suggested $A=12$ as the most appropriate for their 
purpose, as the higher orders did not 
provide any significant improvement.

Alternative discretisations of the Laplacian operator 
are possible, such as the Mehrstellen discretisation 
of Briggs et al. \cite{BRIGGS-BERNHOLC}. This is a 
fourth order discretisation that includes off-diagonal 
terms, but only nearest neighbours to the point of 
interest. It is more costly to compute than the 
standard fourth order formula of equation (\ref{LAPLACIAN_FD}) and 
it is still not clear whether its fourth order 
is sufficient. One may also use FD methods on a 
grid with variable spatial resolution, such as 
that of Modine et al. \cite{ACRES} which is denser 
near the ionic positions. Such a scheme, however, has 
the added overhead of a transformation of the 
Laplacian from Cartesian to curvilinear 
coordinates. In this paper we use only the 
FD scheme of equation (\ref{LAPLACIAN_FD}).

The FD approach has desirable properties, both 
in terms of computational scaling and 
parallelisation. The Laplacian in the FD representation 
is a near-local operator, becoming more 
delocalised with increasing order. Therefore, the 
cost of applying it to $N$ grid points is 
strictly linear (compared to $N\log N$ for 
Fourier transform methods). Also, as a 
result of its near-locality, ideal load balancing 
can be achieved in parallel implementations 
by partitioning the real space grid into 
subregions of equal size and distributing them 
amongst processing elements (PEs) while 
requiring little communication for 
applying the Laplacian at the bordering 
points of the subregions. 

If $N_s$ represents the size of the system, then 
the number of support functions will be 
proportional to $N_s$ and so will the 
total number of grid points in the simulation 
cell, resulting in a total computational 
cost proportional to $N_s^2$ for the 
application of the Laplacian on all support 
functions. More favourable scaling can be achieved 
by predicting the region in space whithin which 
the values of a particular function will be of significant 
magnitude and operating only on this 
region \cite{CHELIKOWSKY2,OOF90}. Linear-scaling 
can be achieved by strictly restricting from 
the outset the support functions to spherical regions 
centred on atoms \cite{GILLAN1}. In this case, the 
cost is $qN_s$ with $q$ being the cost of 
applying the Laplacian on the points of 
a spherical region, which is constant with system size.

FD methods nevertheless have disadvantages that 
do not appear in the plane-wave formalism. Firstly, there 
is no \emph{a priori} way of knowing whether a 
particular order of FD approximation will be sufficient 
to represent a particular support function 
accurately. In addition, while plane-wave methods 
can handle different symmetry groups trivially 
through the reciprocal lattice vectors of the 
simulation cell, real space implementations need to 
consider every symmetry separately and require 
considerable modifications to the code and higher 
computational cost. Briggs et al. \cite{BRIGGS-BERNHOLC} have 
demonstrated this difficulty by performing calculations  
with hexagonal grids while most common applications 
of real space methods in the literature are limited 
to grids of cubic or orthorhombic 
symmetry \cite{CHELIKOWSKY2,MARTIN-MULTI,CHELIKOWSKY1,GILLAN1}.

The computational cost for the calculation of the 
Laplacian of a single support function with the 
FD method scales as $(3A+1)(1+A/D)^3 N_{\mathrm{reg}}$ where 
$N_{\mathrm{reg}}$ is the number of grid points 
within the support region, and $D$ is the number 
of grid points along the support region 
diameter and is proportional to $N_{\mathrm{reg}}^{1/3}$. This 
estimate of cost includes all the nonzero values of 
the Laplacian, which in general occur not only at the 
grid points inside the support region but also at 
points outside, up to a distance of $A/2$ points 
from the region's boundary. It is important to 
include the contribution to the Laplacian from 
outside the support region in the sum of 
equation (\ref{REAL_SUM}) in order to obtain the 
best possible accuracy for a given order $A$ and 
also to ensure the Hermiticity of the discretised 
representation of the Laplacian, $\hat{T}$, and 
hence of the kinetic energy matrix elements $T_{\alpha \beta}$.

\subsection{Localised discrete Fourier transform}

We now present an accurate, linear-scaling method for 
calculating the kinetic energy matrix elements $T_{\alpha\beta}$ 
of equation (\ref{TAB}). We use a mixed space Fourier 
transform approach that is applicable to any Bravais 
lattice symmetry. Fourier transformation is a natural 
method to adopt for this task since in a total-energy 
calculation one computes other terms, such as the 
electron density and the Hartree energy, using reciprocal 
space techniques. This implicitly defines the basis set that 
we use to be plane-waves and for consistency we should 
calculate the kinetic energy using the same basis 
set, i.e.\ using Fourier transform methods. Thus we calculate 
the $\nabla^{2} \phi$ term in reciprocal space, where 
the Laplacian operator is easy to apply, then transform 
the result back to real space and obtain the matrix 
elements $T_{\alpha\beta}$ by summation over grid 
points (\ref{REAL_SUM}). One way to achieve this would 
be to perform a discrete FFT on each support 
function $\phi$, using the periodicity of the entire 
simulation cell. However, unlike the FD algorithm, the 
FFT is not a local operation and the cost of applying 
the Laplacian to all the support functions in this 
way would be proportional to $N_{s}^{2}\log N_{s}$, which 
clearly does not scale linearly with system size.

\begin{figure}

\begin{center}

\centering

\scalebox{0.45}[0.45]{\includegraphics*[0.0cm,0cm][29cm,17cm]{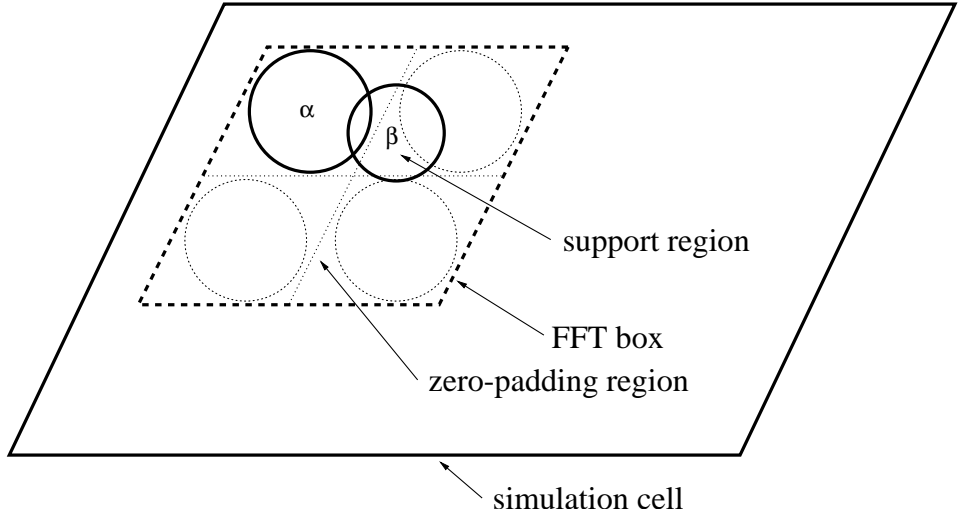} }

\caption{The simulation cell and an FFT box for a pair of overlapping support regions. \label{FFT_BOX} }

\end{center}

\end{figure}

It is possible to overcome this undesirable scaling 
without compromising accuracy by performing the FFT 
over a restricted region of the simulation cell, which 
we call the `FFT box' (figure \ref{FFT_BOX}). Before 
defining the FFT box, there are two points that should 
be noted. Firstly, the operator $\hat{T}$ must be 
Hermitian. This will ensure that the kinetic energy 
matrix elements, $T_{\alpha \beta}$ are Hermitian, and 
hence the eigenvalues real. Secondly, when calculating 
two matrix elements such as $T_{\alpha \beta}$ and 
$T_{\gamma \beta}$, we require the quantity $\hat{T} \phi_{\beta}$ 
in both cases. To be consistent, our method for 
calculating the matrix elements must be such 
that $\hat{T} \phi_{\beta}$ is the same in both 
cases, i.e.\ we require $\hat{T} \phi_{\beta}$ to have 
a unique and consistent representation throughout the 
calculation. It is important that both these conditions 
are satisfied when it comes to optimisation of the support 
functions during a total-energy calculation, and 
we shall return to this point later.

In order to fulfil the above requirements, it can be seen 
that for a given calculation the FFT box must be universal in 
shape and dimensions. As a result, it must be large enough to 
enclose any pair of overlapping support functions within 
the simulation cell. To define a suitable FFT box, we first 
consider a box with the same unit lattice vectors as the 
simulation cell, but of dimensions such that it exactly circumscribes 
the largest support region present in the simulation 
cell. We then define a box that is commensurate with 
this, but with sides that are twice as long (and hence a 
volume eight times as large). This we define to be 
the FFT box. It is clear that this FFT box is large enough 
to enclose any pair of support functions exhibiting any degree of overlap.

To calculate a particular matrix element $T_{\alpha\beta}$ for 
two overlapping support 
functions $\phi_{\alpha}$ and $\phi_{\beta}$, we imagine 
them as being enclosed within 
the FFT box defined above and we treat 
this region of real space as a miniature simulation cell. 
We Fourier transform $\phi_{\beta}$ using the 
periodicity of the FFT box and apply the 
Laplacian at each reciprocal lattice point 
using standard plane-wave techniques \cite{PAYNE_REVIEW}. 
It is then a simple matter of using one more FFT to 
back-transform $\nabla^{2}\phi_{\beta}$ 
to real space and subsequently calculate $T_{\alpha\beta}$ 
by summation over the grid 
points of the FFT box, according to equation (\ref{REAL_SUM}).

The result obtained by this process is equivalent 
to performing a Fourier transform of $\phi_{\beta}$ 
over the whole simulation cell, applying the 
Laplacian and then interpolating to a coarse, but still 
regular, reciprocal space grid with only $N_{box}$ 
points, $N_{box}$ being the number of grid 
points in the FFT box,  before back-transforming to 
real space. This coarse sampling in reciprocal 
space has a negligible influence on the result 
because each support function is strictly localised 
in real space and therefore smooth in reciprocal space.

It is worth noting the implicit approximation that 
we make in calculating the kinetic energy 
in the way prescribed above. In 
general, $\nabla^{2}\phi_{\beta}$ is nonzero outside the support 
region of $\phi_{\beta}$ itself, and it is 
essential to take this into account in the calculations. By 
construction, we neglect contributions to the 
kinetic energy from support functions whose support 
regions do not overlap as we expect them to be 
negligibly small. This approximation may be 
controlled via a single parameter, the FFT box 
size, with respect to which the calculation may be 
converged if necessary. The same approximation 
is of course present in the FD method as well.

We expect certain advantages to the FFT box algorithm 
over FD based methods. Firstly, the FFT box method 
should be more accurate than any FD scheme since it 
takes into account information from every single point 
of the support function and not only locally. However, it 
is still perfectly local as far as parallelisation 
is concerned since we only deal with the points within 
a single FFT box each time, and this constitutes a 
very small region of the simulation cell. The parallelisation 
strategy in this case would still consist of 
partitioning the real space grid of the 
simulation cell into subregions of equal size 
and distributing them amongst PEs. Then, FFTs local 
to each PE are performed on FFT boxes enclosing pairs 
of overlapping support regions belonging completely to 
the simulation cell subregion of the given PE. For 
pairs of overlapping support regions containing grid points 
common to the subregion  of more than one PE, the pair 
would have to be attributed to one PE and copied as a 
whole to it for the local FFT to proceed. This would 
involve some communication overhead, as in the FD case 
for pairs of overlapping support functions with points in 
more than one subregion. Another important advantage 
of the FFT box method is that it is applicable, without 
any modification, to regular grids of any 
Bravais lattice symmetry. This is not true of FD methods.

The number of grid points in a cubic FFT box 
is $N_{box}$ (which is related to $N_{\mathrm{reg}}$ by $N_{box}=
8 \times 6 N_{\mathrm{reg}}/\pi \simeq 15.3 N_{\mathrm{reg}}$). Therefore 
the computational cost of applying the FFT method 
to a single support function in such an FFT box is 
$2 N_{box} \log N_{box}$, and thus for all support functions
the cost is proportional to $2 N_s N_{box} \log N_{box}$,
where $N_{box}$ is independent of $N_s$. In 
other words the cost scales linearly 
with the number of atoms in the system.

\section{Tests and discussion}

\begin{figure}

\vspace{-6em}

\begin{center}

\scalebox{0.52}{\includegraphics*[0cm,0.0cm][27cm,19cm]{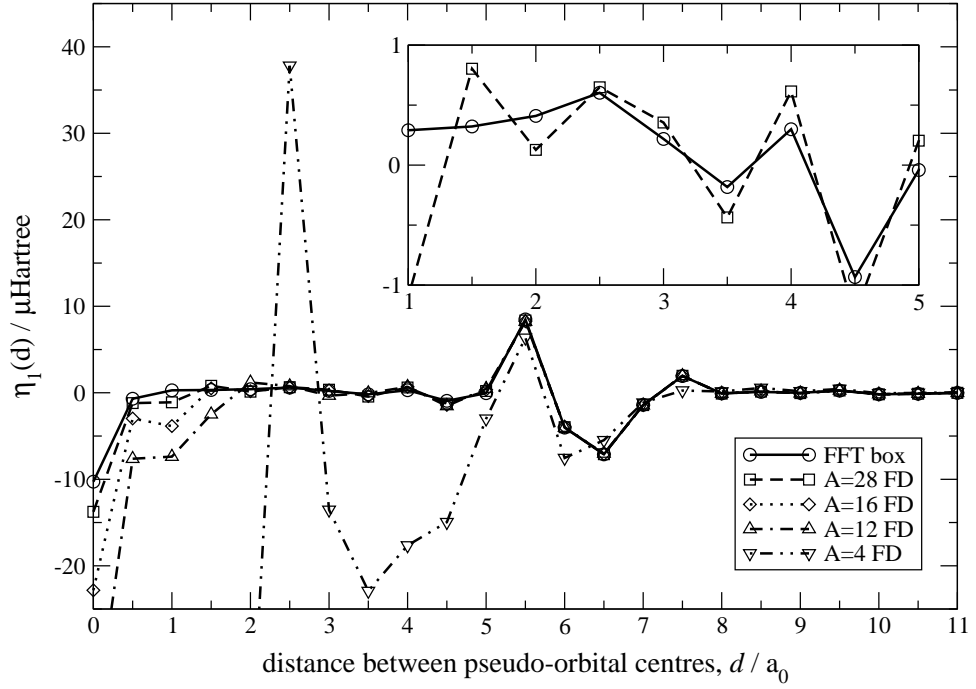} }

\scalebox{0.52}{\includegraphics*[0cm,0.0cm][27cm,19cm]{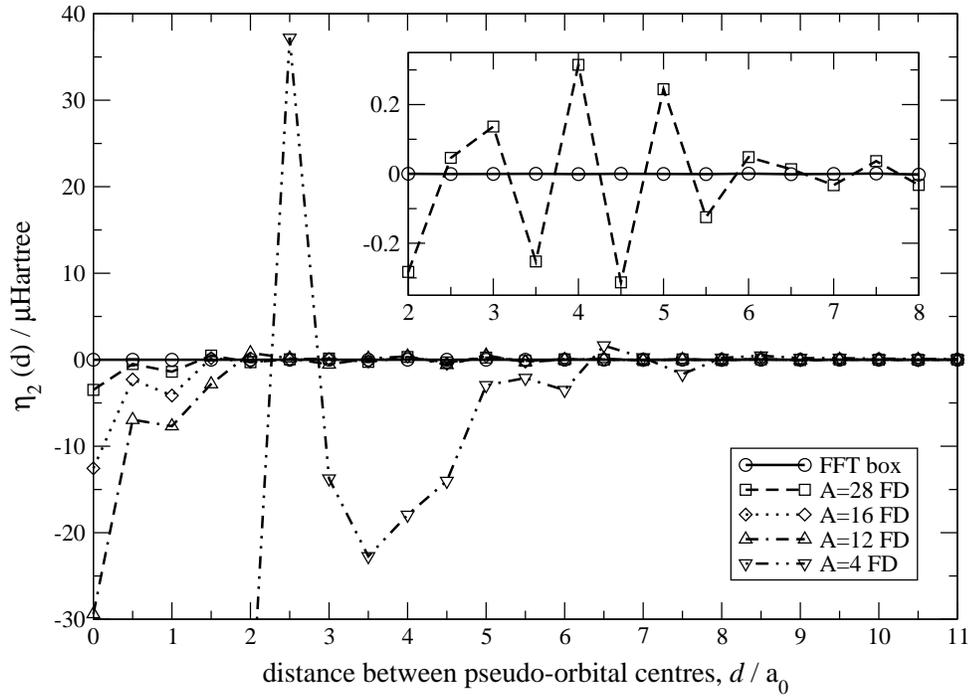} }

\vspace{-2em}

\caption{Top panel: $\eta_{1}(d)$ for a carbon 2s valence 
pseudo-orbital in a spherical support region with a radius 
of $6.0 \: a_{0}$. Bottom panel: $\eta_{2}(d)$ for the 
same pseudo-orbital. The insets show a magnification 
of the plots for \mbox{A = 28} FD and the FFT box method. \label{ERRORS}}

\end{center}

\end{figure}

We have performed tests of the FD and FFT box 
methods for calculating the kinetic energy 
of localised functions. Choosing a particular type 
of support function $\phi$ with spherical symmetry, placing 
one at $\mathbf{R}_\alpha$ and another at $\mathbf{R}_\beta$, we 
rewrite the integral of equation (\ref{TAB}) as
\begin{equation}
T(|\mathbf{R}_\alpha-\mathbf{R}_\beta|) = -\frac{1}{2}\int \phi^*(\mathbf{r}-\mathbf{R}_\alpha) 
\nabla^2 \phi(\mathbf{r}-\mathbf{R}_\beta) \mathrm{d} \mathbf{r} \ .
\end{equation}
For our first test we calculate the following quantity 
as a function of the distance $d$ between the 
centres $\mathbf{R}_\alpha$ and $\mathbf{R}_\beta$
\begin{equation}
\eta_{1}(d)=T_{ap}(d)-T_{ex}(d) \ ,
\end{equation}
where $T_{ex}(d)$ is the exact value of the integral 
in the continuous representation of the support functions 
and $T_{ap}(d)$ is its approximation on the real space 
grid, either by FD or the FFT box method. We 
chose $\phi(\mathbf{r})$ to be a 2s valence pseudo-orbital 
for a carbon atom, generated using an atomic norm-conserving 
carbon pseudopotential \cite{TROULLIER1} within the local 
density approximation. The pseudo-orbital is confined in a 
spherical region of radius $6.0 \: a_{0}$, and vanishes 
exactly at the region boundary \cite{SANKEY_FIREBALLS}. It is 
initially generated as a linear combination of spherical 
Bessel functions, which are the energy eigenfunctions of 
a free electron inside a spherical box. Our functions are 
limited up to an energy of 800 eV, resulting in a combination 
of fourteen Bessel functions. The formula for calculating 
kinetic energy integrals between Bessel functions is 
known \cite{HAYNES97} and we used it to obtain $T_{ex}(d)$ for 
our valence pseudo-orbital. We then calculated $\eta_{1}(d)$ with 
a grid spacing of $0.4 \: a_0$ (corresponding to a plane-wave 
cut-off of 839 eV) in an orthorhombic simulation cell, as 
we are restricted to do so by the FD method. With these 
parameters $N_{box}$ is $60^{3}$, and hence it is trivial 
to perform the FFT of one support function on a single 
node. $\eta_{1}(d)$ is plotted for the FFT box method 
and for various orders of the FD method in the 
top graph of figure \ref{ERRORS}.

It can be seen that low order FD methods are inaccurate as 
compared to the FFT box method, and only when order 28 FD is 
used does the accuracy approach that of the FFT box 
method. The \mbox{A = 12 FD} scheme, the highest order that 
has been used in practice for calculations \cite{CHELIKOWSKY1}, gives 
an error of $-3.97 \times 10^{-5} \: \mbox{Hartree}$ at $d=0$ as 
compared to $1.027 \times 10^{-5} \: \mbox{Hartree}$ for 
the FFT box method. The feature that occurs in the top graph 
of figure \ref{ERRORS}, between $d=5 \: a_0$ and $d=7 \: a_0$, is 
an artefact of the behaviour of our pseudo-orbitals at the 
support region boundaries where they vanish exactly, but 
with a finite first derivative. This causes an enhanced 
error in all the methods when the edge of one support 
function falls on the centre of another.

The error in the FFT box method is small, yet 
non-zero, and we attribute this to the inherent discretisation 
error associated with representing functions that are not 
bandwidth limited on a discrete real space 
grid. Convergence to the exact result is observed 
as the grid spacing is reduced, as expected.

As our next comparison of the FFT box and FD methods we 
used the same pseudo-orbitals as before, but considered the quantity 
\begin{equation}
\eta_{2}(d)=T_{ap}(d)-T_{PW}(d)
\end{equation}
as the measure of the error, where $T_{PW}$ is the result 
obtained by Fourier transforming the support functions 
using the periodicity of the entire simulation cell. One 
may think of $T_{PW}$ as being the result that would 
be obtained from a plane-wave code: the support 
functions may be considered as generalised Wannier 
functions. Calculating the kinetic energy integrals by 
performing a discrete Fourier transform on the support 
functions over the entire simulation 
cell (an $O(N_{s}^{2}\log N_{s})$ process for all support functions) 
is equivalent to summing the contributions to the 
kinetic energy from all of the plane-waves up to 
the cut-off energy determined by the grid 
spacing. Thus our FFT box method can be viewed 
as equivalent to a plane-wave method that uses a 
contracted basis set (i.e.\ a coarse sampling in reciprocal space). In 
some ways $\eta_{2}(d)$ is a better measure of the relative 
accuracy of the FD and FFT box methods as our goal is to 
converge to the `exact' result as would be obtained 
using a plane-wave basis set over the entire 
simulation cell. $\eta_{2}(d)$ is plotted in the bottom graph of figure \ref{ERRORS}.

$T_{PW}$ was computed using a cell that contained 256 grid 
points in each dimension. Increasing the cell size further had 
no effect on $T_{PW}$ up to the eleventh decimal 
place ($10^{-11} \: \mbox{Hartree}$). The plots show that 
the FFT box method performs significantly better than 
all orders of FD that were tested. For example at $d=0$ the 
error for \mbox{A = 28 FD} is $-3.49 \times 10^{-6} \: \mbox{Hartree}$ as 
compared to $-1.09 \times 10^{-9} \: \mbox{Hartree}$ for 
the FFT box method. The fact that the FFT box error is 
so small shows that coarse sampling in reciprocal space 
has little effect on accuracy, as one would expect 
for functions localised in real space.

Our implementation can produce similar FFT box results 
to the above in regular grids of arbitrary 
symmetry (non-orthogonal lattice vectors) as long as 
we include roughly the same number of grid points in 
the support region sphere. As we described earlier the 
application of the FD method to grids without 
orthorhombic symmetry is not straightforward.

Furthermore, in our implementation the kinetic energy 
matrix elements $T_{\alpha\beta}$ for both the FFT box 
method and the FD method (of any order) are Hermitian 
to machine precision. This is a direct consequence 
of $\hat{T}$, our representation of the Laplacian 
operator $\nabla^{2}$ on the grid, being Hermitian. As 
mentioned earlier, this is an important point. The 
matrix elements $T_{\alpha \beta}$ may always 
be made Hermitian by construction without $\hat{T}$ itself 
being an Hermitian operator. This would ensure 
real eigenvalues, as is required. However, when 
it comes to optimisation of the support functions 
during a total-energy calculation, we require 
the derivative of the kinetic energy with respect 
to the support function values \cite{GILLAN1}:
\begin{equation}
\frac{\partial E_T}{\partial \phi^{*}_{\alpha} (\mathbf{r}_i) } = 
-\frac{1}{2} \sum_\beta K^{\beta\alpha}\hat{T} 
\phi_\beta (\mathbf{r}_i ) \ , \: \frac{\partial E_T}{\partial \phi_{\alpha} (\mathbf{r}_i) } 
= -\frac{1}{2} \sum_\beta K^{\alpha\beta}\hat{T} \phi^{*}_{\beta} (\mathbf{r}_i ) \ , \label{GRADIENT}
\end{equation}
where the $\mathbf{r}_i$ are grid points belonging 
to the support region of $\phi_\alpha$. These relations 
both hold only if $\hat{T}$ is an Hermitian operator, and 
support function optimisation can only be performed 
in a consistent manner if there is one unique representation 
of $\hat{T} \phi_{\beta}$ for each support 
function $\phi_{\beta}$. It is also worth noting that the 
evaluation of these derivatives is the reason why we prefer 
to perform the sum of equation (\ref{REAL_SUM}) for 
the FFT box method in real space, rather than in its 
equivalent form in reciprocal space. Applying the FFT 
box method in reciprocal space would be no more costly 
as far as integral evaluation is concerned but we would 
require an extra FFT per support function for the subsequent 
evaluation of equation (\ref{GRADIENT}).

For all the methods we describe in this paper we observe 
variation in the values of the kinetic energy integrals when 
we translate the system of the two support functions 
with respect to the real space grid. This is to 
be expected as the discrete representation of the 
support functions changes with the position of the 
support region with respect to the grid. Such 
variations may have undesirable consequences when it 
comes to calculating the forces on the atoms. In 
FFT terminology, they result from irregular 
aliasing of the high frequency components of our 
support functions as they are translated in real 
space. Ideally, in order to avoid this effect, the 
reciprocal representation of the support functions 
should contain frequency components only up to the 
maximum frequency that corresponds to our grid 
spacing, in other words it should be strictly localised 
in reciprocal space. Unfortunately this constraint 
is not simultaneously compatible with strict real 
space localisation. It should be possible however 
to achieve a compromise, thus controlling the 
translation error by making it smaller than some 
threshold. Such a compromise should involve an 
increase in the support region radii of our functions 
by a small factor. This situation is similar 
to the calculation of the integrals of the nonlocal 
projectors of pseudopotentials in real space 
with the method of King-Smith et al. \cite{KINGSMITH-REAL} which 
requires an increase of the core radii by a 
factor of 1.5 to 2. For example, if we consider 
two carbon valence pseudo-orbitals of support 
radius $6.0 \: a_0$ and with $d=5.0 \: a_0$ and translate 
them both in a certain lattice vector direction 
over a full grid spacing, the maximum variation 
in the value of the integral  with the FFT box 
method is  $8.28 \times 10^{-6}$ Hartree. If we 
then do the same with carbon pseudo-orbitals generated 
with precisely the same parameters but instead 
with a support radius of $10.0 \: a_0$, the maximum 
variation with respect to translation 
is reduced to $2.05 \times 10^{-8}$ Hartree.

\section{Conclusions}

In conclusion, we have presented a new and easy 
to implement method for calculating kinetic energy 
matrix elements of localised functions represented on 
a regular real space grid. This FFT box method is 
based on a mixed real space -- reciprocal space 
approach. We use well established FFT algorithms to 
calculate the action of the Laplacian operator on 
localised support functions, whilst maintaining 
linear-scaling with system size and near locality 
of the operation. This makes our FFT box method suitable 
for implementation in the order-N code that we are 
developing. We have performed tests of the FFT box 
method and various orders of FD. Comparing to the 
exact integrals of the continuous representation, we 
have demonstrated that our approach is more accurate 
than low order FD approximations and only 
when \mbox{A = 28} FD is used does the accuracy 
become comparable to that of the FFT box method. We 
have also highlighted the connection between the 
FFT box method and plane-wave methods and shown 
that our approach is up to three orders of magnitude 
more accurate than \mbox{A = 28} FD when compared 
to the `exact' result within the plane-wave basis 
set of the entire simulation cell. Furthermore, our 
approach for calculating the kinetic energy is 
consistent with the way in which other quantities 
in a total-energy calculation, such as the 
electron density and the Hartree energy, are 
computed as these are also calculated using 
reciprocal space techniques. Finally, we also note 
that our FFT box method is more versatile 
than FD as it is applicable to real space 
grids based on any lattice symmetry whereas 
FD schemes are usually only applied to orthorhombic grids.

\section{Acknowledgements}
C.-K. S. would like to thank the EPSRC (grant number GR/M75525) 
for postdoctoral research funding. A. A. M. would like to thank 
the EPSRC for a Ph.D. studentship. P. D. H. would like to 
thank Magdalene College, Cambridge for a Reseach 
Fellowship. C. J. P. would like to thank the 
EPSRC (grant number GR/N18567) for a ROPA.

\end{document}